\documentclass[twocolumn]{aastex631}

\newcommand{\Rom}[1]{\uppercase\expandafter{\romannumeral #1\relax}}

\makeatother

\shorttitle{Two-Component Jet for GRB 221009A}
\shortauthors{Zheng et al.}

\graphicspath{{./}{figures/}}

\begin{document}

\title{A Narrow Uniform Core with a Wide Structured Wing: Modeling the TeV and Multi-wavelength Afterglows of GRB 221009A}

\author[0000-0001-5751-633X]{Jian-He Zheng}
\affiliation{School of Astronomy and Space Science, Nanjing University, Nanjing 210023, People’s Republic of China}
\affiliation{Key laboratory of Modern Astronomy and Astrophysics (Nanjing University), \\
Ministry of Education, Nanjing 210023, People’s Republic of China}

\author[0000-0002-5881-335X]{Xiang-Yu Wang}
\affiliation{School of Astronomy and Space Science, Nanjing University, Nanjing 210023, People’s Republic of China}
\affiliation{Key laboratory of Modern Astronomy and Astrophysics (Nanjing University), \\
Ministry of Education, Nanjing 210023, People’s Republic of China}
\email{xywang@nju.edu.cn}

\author[0000-0003-1576-0961]{Ruo-Yu Liu}
\affiliation{School of Astronomy and Space Science, Nanjing University, Nanjing 210023, People’s Republic of China}
\affiliation{Key laboratory of Modern Astronomy and Astrophysics (Nanjing University), \\
Ministry of Education, Nanjing 210023, People’s Republic of China}

\author[0000-0002-9725-2524]{Bing Zhang}
\affiliation{Nevada Center for Astrophysics, University of Nevada Las Vegas, Las Vegas, NV 89154, USA}
\affiliation{Department of Physics and Astronomy, University of Nevada Las Vegas, Las Vegas, NV 89154, USA}



\begin{abstract}

The TeV afterglow of the BOAT GRB 221009A was interpreted as arising from a narrow jet while the radio to X-ray afterglows were interpreted as arising from a wide structured jet. However, there is no model explaining the TeV and lower-energy multi-wavelength afterglows simultaneously.  We here investigate a two-component jet model, including a narrow uniform core with  a wide structured wing, to explain both the  multi-wavelength afterglows that last up to 100 days.  We find that to explain the early TeV afterglow with the inverse-Compton process, we need a  circum-burst density higher than $\ga 0.1{\rm cm^{-3}}$, while the radio afterglow and the H.E.S.S. upper limit  combine to constrain the  density to be lower at larger radii. Thus, a decreasing density profile with radius is favored. Considering that the rising TeV light curve during the afterglow onset favors a constant-density medium, we invoke a stratified density profile, including a constant-density profile  at small radii and a wind density profile at large radii. We find that the two-component jet model with such a stratified density profile can explain the TeV, X-ray and optical afterglows of GRB 221009A, although the radio fluxes exceed the observed ones by a factor of two at later epochs. The discrepancy in the radio afterglow could be resolved by invoking some non-standard assumption about the microphysics of afterglow shocks. The total kinetic energy of the two components in our model is $\lesssim 10^{52}{\rm erg}$, significantly smaller than that in the single structured jet models.

\end{abstract}



\section{Introduction} \label{sec:intro}
GRB 221009A  is the brightest-of-all-time  (BOAT) GRB \citep{2023ApJ...946L..31B}. Due to its enormous energy ($E_{\gamma,\rm iso}=1.5\times10^{55}$ erg) \citep{GECAM2023arXiv230301203A,Lesage2023ApJ...952L..42L} and
proximity ($z=0.151$) \citep{Castro2022GCN.32686....1C}, GRB 221009A is an exceptionally rare event \citep{Connor2023SciA....9I1405O,2023ApJ...946L..31B,Williams2023ApJ...946L..24W}. The Large High Altitude Air Shower Observatory (LHAASO) observed GRB 221009A at the earliest epoch, covering both the prompt emission phase and the early afterglow in TeV band, and revealed the onset of afterglow emission in the TeV band for the first time\citep{LHAASO2023Sci...380.1390L}. 

In addition, the temporal slope of the TeV light curve in the decaying phase steepens from $\alpha=-1.12^{+0.01}_{-0.01}$ to $\alpha=-2.21^{+0.30}_{-0.83}$ at $t_{\rm b}=670^{+230}_{-110}{\rm s}$ after the afterglow onset time $T^{*}=T_0+226 {\rm s}$, where $T_0$ is the trigger time of this GRB\footnote{We note $F_{\nu}\propto\nu^{\beta}t^{\alpha}$ throughout this paper}, indicating that the opening angle of the jet of GRB 221009A is only $\sim0.8^{\circ}$\citep{LHAASO2023Sci...380.1390L}. This reduces the beaming-corrected energy in gamma-rays to a level of $10^{50}-10^{51}{\rm erg}$ for GRB 221009A, which agrees  with the  beaming-corrected gamma-ray energy release in other GRBs \citep{Frail2001ApJ...562L..55F}. 

It has been suggested that jets of GRBs may not be uniform, but characterized by a significant anisotropy of the angular distribution of the fireball energy around the axis \citep{Rossi2002MNRAS.332..945R,Zhang2002ApJ...571..876Z}.  
In the assumption of a structured jet model, the small half opening angle of GRB 221009A implies a narrow core component, which is only responsible for the early afterglow emission before $10^4{\rm s}$ \citep{LHAASO2023Sci...380.1390L}, while the  late-time ($>10^4$s) afterglow emission  may require other jet components. 

The X-ray afterglow of GRB 221009A  features an initial power-law decay index of $\alpha_1=-1.52\pm0.01$,
steepening to $\alpha_2=-1.66\pm0.01$ after $0.82\pm0.07$ days after the trigger, which  is not consistent with standard predictions for the emission from a top-hat jet \citep{Connor2023SciA....9I1405O,Williams2023ApJ...946L..24W}.  \cite{Connor2023SciA....9I1405O}  interpreted the X-ray afterglow as due to a structured jet expanding into a constant-density medium, where the jet is composed by an inner component of angular size $\theta_b$ with a shallow energy profile $dE/d\Omega \propto \theta^{-a_1}$ and slightly steeper lateral structure at $\theta>\theta_{\rm b}$ with $dE/d\Omega \propto \theta^{-a_2}$ ($a_2>a_1$).  \citet{Gill2023MNRAS.524L..78G}  also explored a  structured jet model  to explain the multi-wavelength afterglow of  GRB 221009A, but assuming a wind density profile for the surrounding medium. In both models,  the early radio emission is
attributed to electrons accelerated by the reverse shock, while the optical and X-ray afterglows, as well as the late radio afterglow, arise from the forward shock.
A two-component jet model consisting of a narrow top-hat jet and a broader top-hat jet has been proposed to explain the multi-wavelength data from radio to GeV afterglows of GRB 221009A \citep{Sato2023MNRAS.522L..56S}.  Besides the two-component jet models, various other types of models were also proposed to explain the multi-wavelength data of GRB 221009A \citep{Ren2023ApJ...947...53R,2023ApJ...947L..14Z}.

The above models, however, did not take into account the early TeV data observed by LHAASO, which were not available at that time.  In this work, we  explore  a two-component jet model, including a narrow top-hat  jet (core)  and a wider wing with an angular structure,  to explain both the TeV afterglow measured by LHAASO and lower energy multi-wavelength afterglows of GRB 221009A.

\section{The set-up of a Two-component jet model}
The afterglow of GRB 221109A exhibits two breaks in light curves. The early sharp break in the TeV band is consistent with a jet break from a top hat jet \citep{LHAASO2023Sci...380.1390L}, while the later shallow break could be due to the change of the angular profile $dE/d\Omega\propto\theta^{-a}$ of a structured jet \citep{Connor2023SciA....9I1405O,Gill2023MNRAS.524L..78G}. This complex behaviour suggests a jet composing of a inner, narrow top-hat  core component and a outer, wide wing component with an angular structure, which can be described by \citep{Zhang&Wang2023}

\begin{eqnarray}
    \epsilon \equiv \frac{dE}{d\Omega} = \left \{
      \begin{array}{ll}
        \epsilon_{\rm I}, & \theta < \theta_{\rm j}, \\
        \epsilon_{\rm II} f(\theta), & \theta_{\rm j} < \theta < \Theta,
      \end{array}
      \right.
\label{eq:epsilon-structure}
\end{eqnarray}
where $\theta_{\rm j}$ is the opening angle of the narrow core, $\Theta$ is the maximum angle of the structured wing, and $f(\theta)$ is the structure function of the wing. The isotropic energy of the wing could be much smaller than that of the core ($\epsilon_{\rm\Rom{1}} \gg \epsilon_{\rm\Rom{2}}$), which can only be determined by the afterglow data. We assume a smooth broken power-law function for the structure of the wing, as given by \citep{Granot2003ApJ...591.1086G}
\begin{equation}
    f(\theta)=
    {{{{\left[ {{{\left( {\frac{\theta }{{{\theta _{\rm c,w}}}}} \right)}^{2{a_1}}} + {{\left( {\frac{\theta }{{{\theta _{\rm c,w}}}}} \right)}^{2{a_2}}}} \right]}^{-1/2}}}},
\end{equation}
where $\theta_{\rm c,w}$ is the  transition angle from a shallow angular profile to a steeper ($a_2>a_1$) angular profile of the wide wing. 
The structured function declines as $f(\theta)\propto\theta^{-a_1}$ from $\theta_{\rm j}$ to  $\theta_{\rm c,w}$ and then transfers into $f(\theta)\propto\theta^{-a_2}$ after $\theta_{\rm c,w}$. The shallow jet break happens when the observers see the edge of  $\theta_{\rm c,w}$.

Correspondingly, one could also define a jet structure of the angle-dependent initial Lorentz factor, i.e.
\begin{eqnarray}
    \Gamma_0(\theta) = \left \{
      \begin{array}{ll}
        \Gamma_{\rm\Rom{1},0}, & \theta < \theta_{\rm j} \\
        \Gamma_{\rm\Rom{2},0} g(\theta), & \theta_{\rm j} < \theta < \Theta,
      \end{array} 
      \right.
\label{eq:Gamma-structure}
\end{eqnarray}    
where $\Gamma_{\rm\Rom{1},0} \gg \Gamma_{\rm\Rom{2},0}$ and $g(\theta)$ is the structure function of the Lorentz factor profile. 

The narrow jet core could be a Poynting-flux dominated jet \citep{Dai2023ApJ...957L..32D,Yang2023ApJ...947L..11Y} and the wide wing could be matter-dominated, as argued in \cite{Zhang&Wang2023}.

\section{A Stratified Density Profile} 
\label{sec:}
The slopes of afterglows produced by forward shocks depend on the density profile of the circum-burst medium. The  density profile of the circum-burst medium is usually described by $n(R)\propto R^{-k}$, where $k=0$ corresponds to a homogeneous medium, while $k=2$ corresponds to a stellar wind from the GRB progenitor \citep{Dai1998MNRAS.298...87D,Chevalier2000ApJ...536..195C,Panaitescu2000ApJ...543...66P}.
In the wind case,  for a constant  mass loss rate $\dot{M}$ and wind velocity $v_{\rm w}$, one has $n=A r^{-2}$, where $A=3\times10^{35}A_{\star}{\rm cm^{-1}}$,  scaled to $A_{\star}=(\dot{M}/10^{-5}M_\odot {\rm yr^{-1}})(v_w/10^3 {\rm Km s^{{-1}}})^{-1}$. 

The  TeV light curve of GRB 221009A rises with a slope of $\alpha = 1.82_{-0.18}^{+0.21}$ before the peak \citep{LHAASO2023Sci...380.1390L}. This rising phase is interpreted as the onset of the TeV afterglow, where the ejecta is coasting before deceleration.
During the coasting phase, the bulk Lorentz factor of the afterglow shock is roughly a constant. The flux resulted from the synchrotron self-Compton process (SSC) rises with a slope as $\alpha=\frac{8-(p+2)k}{4}$  if the observed frequency lies above the peak frequency of the SSC spectrum \citep{LHAASO2023Sci...380.1390L}. The rising slope $\alpha = 1.82_{-0.18}^{+0.21}$ of the TeV light curve is  consistent with the constant-density medium (i.e., $k=0$). Note that this conclusion applies only to the small radius where the early ($t<18$s) TeV emission is produced. 

The density at a larger distance could be different and can only be constrained by late-time multi-wavelength data.
Fermi-LAT  measurement of GeV emission at $T_{0}+20000$s suggests a low density at large radii, as we show below. 
The High Energy Stereoscopic System (H.E.S.S.) began observations 2.5 days after the trigger, yielding an upper limit on the TeV flux \citep{HESS2023ApJ...946L..27A}. This upper limit implies a small TeV to keV flux ratio at 2.5 days, which also supports a low density at larger radii.

\subsection{Lower Limit on the Density at Small Radii }
We assume that the TeV emissions observed by LHAASO are produced by the afterglow synchrotron self-Compton (SSC) process. Then we can use the TeV flux to constrain the density of the surrounding medium.  The SSC flux without considering the Klein-Nishina (KN) effect can be expressed as \citep{Sari2001ApJ...548..787S}. 
\begin{equation}
F_{\nu}= \left\{
\begin{array}{lll}
F_m^{\rm IC} (\frac{\nu}{\nu_m^{\rm IC}})^{\frac{1}{3}} ,  \,\,\,\, &\nu<\nu_m^{\rm IC} \\
F_m^{\rm IC} \left(\frac{\nu}{\nu_m^{\rm IC}}\right)^{-\frac{p-1}{2}} , \,\,\,\, &\nu_m^{\rm IC}<\nu<\nu_c^{\rm IC}  \\
F_m^{\rm IC} (\frac{\nu_c^{\rm IC}}{\nu_m^{\rm IC}})^{\frac{1}{2}}(\frac{\nu}{\nu_m^{\rm IC}})^{-\frac{p}{2}}  , \,\,\,\, &\nu>\nu_c^{\rm IC}
\end{array}
\right.
\end{equation} 
where $F^{\rm IC}_{\rm m}=\tau_{\rm IC}F_{\rm \nu,max}$ is the peak flux density for SSC, which scales as $F^{\rm IC}_{\rm m}\propto n^{5/4}_{0}$ in a constant-density medium,  $p$ is the spectral index for shock-accelerated elections ($dN_{\rm e}/d\gamma^{\prime}_{\rm e}\propto{\gamma^{\prime}_{\rm e}}^{-p}$), $\nu^{\rm IC}_{\rm m}$ and $\nu^{\rm IC}_{\rm c}$ are the characteristic inverse Compton frequency for minimal Lorentz factor $\gamma^{\prime}_{\rm m}$ and cooling Lorentz factor $\gamma^{\prime}_{\rm c}$, respectively (see the Appendix \ref{app:ana} for more details).

We assume that only a fraction $\xi_{\rm e}$ of shock-heated electrons are accelerated into a power-law form.
The minimum break frequency in the SSC spectra is
\begin{equation}
    \label{numIC}
    h\nu^{\rm IC}_{\rm m}=1{\rm GeV}E^{3/4}_{\rm \Rom{1},iso,55}\epsilon^{4}_{\rm e,-1.5}\xi^{-4}_{\rm e,0}\epsilon^{1/2}_{\rm B,-4}n^{-1/4}_{0,-0.5}t^{-9/4}_{1.5},
\end{equation}
where $n_0$ is the number density of the circum-burst medium, $\epsilon_{\rm e}$ is the energy equipartition factor for non-thermal electrons, $\epsilon_{\rm B}$ is the energy equipartition factors for the  magnetic field (we use $p=2.3$ to derive the coefficient in the equation). Hereafter,  we adopt the convention that subscript numbers $x$ indicate normalisation by $10^x$ in cgs units.

LHAASO observations show the spectral index in 0.2-7 TeV is $\beta\simeq -1.4$, suggesting $\nu^{\rm IC}_{\rm m}\la200$GeV. 
This imposes a limit on  $k_{\rm e}\equiv\epsilon_{\rm e}/\xi_{\rm e}$,
\begin{equation}
    k_{\rm e}\leq0.15E^{-3/16}_{\rm\Rom{1}, iso,55}\epsilon^{-1/8}_{\rm B,-4}n^{1/16}_{0,-0.5}.
\end{equation}
In the spectral region $\nu\geq\nu^{\rm IC}_{\rm m}$, 
the flux density at $h\nu=300 {\rm GeV}$ at 30s is given by
\begin{equation}
    \label{SSCflux}
    F_{\nu}({\rm 300GeV})=0.005\mu{\rm Jy}E^{\frac{7+3p}{8}}_{\rm \Rom{1},iso,55}k^{2(p-1)}_{\rm e,-1.5}\xi_{\rm e,0}n^{\frac{11-p}{8}}_{0,-0.5}\epsilon^{\frac{p+1}{4}}_{\rm B,-4}.
\end{equation}
Since the KN effect will only suppress the SSC process, the flux given by Equation (\ref{SSCflux}) can be regarded as  an upper limit for the observed flux density, which is $F_{\rm \nu,obs}({\rm 300GeV})=0.0052\mu{\rm Jy}$ at $T^{*}+30$s \citep{LHAASO2023Sci...380.1390L}. Note that, if  $\nu^{\rm IC}_{\rm c}<300{\rm GeV}$, the spectrum is steeper than $F_{\nu}\propto\nu^{(1-p)/2}$, then the inequality $F_{\nu}({\rm 300GeV})\geq F_{\rm \nu,obs}({\rm 300GeV})$ is still applicable.

Considering that the afterglow synchrotron flux should not be greater than the observed flux by Fermi/LAT at 100 MeV  at $T^{*}+110$s, we obtain $\xi_{\rm e,0}k^{p-1}_{\rm e,-1.5}\leq E^{-\frac{(p+2)}{4}}_{\rm iso,55}\epsilon^{-\frac{(p-2)}{4}}_{\rm B,-4}$ \citep{LHAASO2023Sci...380.1390L}. 
Utilizing the above maximum value of $\xi_{\rm e}k^{p-1}_{\rm e}$ and $k_{\rm e}$, we obtain a lower limit on the circum-burst density at the shock radius corresponding to $t=T^{*}+30$s,
\begin{equation}
    \label{highden}
    n_{0}\geq0.1{\rm cm^{-3}}E^{-\frac{9-p}{21-p}}_{\rm \Rom{1},iso,55}\epsilon^{-\frac{2(7-p)}{21-p}}_{\rm B,-4}.
\end{equation}
Below we will show that  the radio afterglow, GeV emission observed by Fermi/LAT and the H.E.S.S. TeV upper limit   combine  to constrain the circum-burst  density to be lower than Equation (\ref{highden}) at larger radii.

\subsection{Upper Limit on the Density at Large Radii}

The observed flux  at 1 GeV is $\mathcal{F}_{\rm obs,GeV}\sim2\times10^{-10}{\rm ergcm^{-2}s^{-1}}$ 
at $T_0+20000$s \citep{Liu2023ApJ...943L...2L} and the flux at 1 keV at the same time is  $\mathcal{F}_{\rm obs,keV}\sim10^{-9}{\rm ergcm^{-2}s^{-1}}$. Considering that the GeV flux of the afterglow is contributed by both the SSC emission and synchrotron emission while the flux at 1 keV is fully contributed by the synchrotron emission, the flux ratio between the SSC component and the synchrotron component  should be smaller than 0.2, i.e., $\mathcal{F}^{\rm IC}_{\rm GeV}/\mathcal{F}^{\rm syn}_{\rm keV}\leq0.2$. Using this flux ratio, we can obtain an upper limit on the circum-burst density, in combination with the radio afterglow data.

The characteristic cooling break in the SSC spectrum are  
$h\nu^{\rm IC}_{\rm c}=100{\rm PeV}E^{-5/4}_{\rm\Rom{2},iso,54}\epsilon^{-7/2}_{\rm B,-4}n^{-9/4}_{0,-0.5}t^{-1/4}_{4.3}$ while the KN peak  $E^{\rm IC}_{\rm c,KN}=0.26{\rm TeV}E^{-1/4}_{\rm\Rom{2},iso,54}\epsilon^{-1}_{\rm B,-4}n^{-3/4}_{0,-0.5}t^{-1/4}_{4.3}$ is around TeV. So the GeV band probably lies between $\nu^{\rm IC}_{\rm m}$ and $\nu^{\rm IC}_{\rm c}$ at $T_0+20000$s. The spectral index of X-ray emission measured by Swift XRT is $\beta=-0.78\pm{0.011}$ at $T_{0}+26000$s,  consistent with the slow cooling phase. 
With the spectra regimes of $\nu_{\rm m}<\nu_{\rm keV}<\nu_{\rm c}$ and $\nu^{\rm IC}_{\rm m}<\nu_{\rm GeV}<\nu^{\rm IC}_{\rm c}$, the expected ratio between  the SSC flux at 1 GeV and the synchrotron flux at 1 keV is
\begin{equation}
    \label{ratio}
    \frac{\mathcal{F}^{\rm IC}_{\rm GeV}}{\mathcal{F}^{\rm syn}_{\rm keV}}=0.019E^{\frac{p+1}{8}}_{\rm\Rom{2},iso,54}k^{p-1}_{\rm e,-1.5}n^{\frac{7-p}{8}}_{0,-0.5}.
\end{equation}

\begin{figure}
    \centering
    \includegraphics[width=0.47\textwidth]{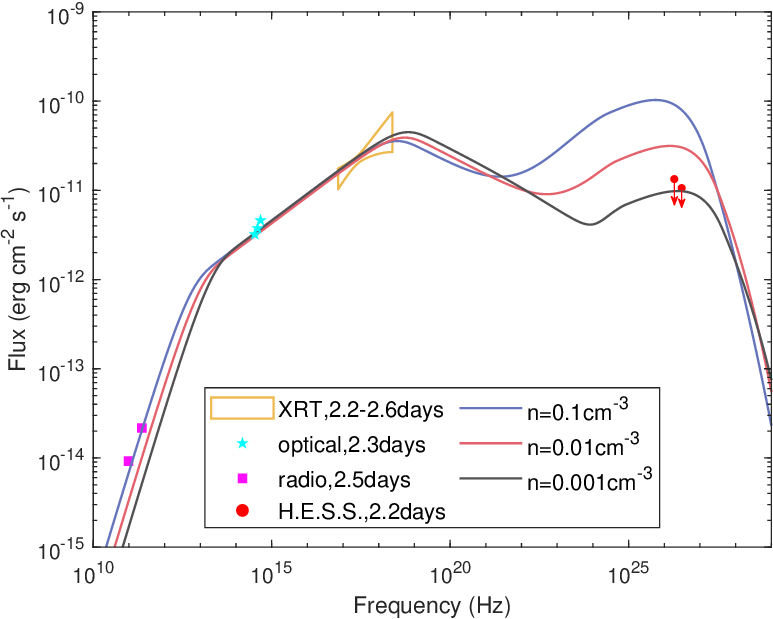}
    \caption{Spectra energy distribution of the afterglow emission at $T_0+$ 2.5 days. The red points denote the 95$\%$ C.L. upper limits from H.E.S.S. Collaborations.  The blue, red and black lines represent the afterglow models with density $n_0=0.1{\rm cm^{-3}}$, $n_0=0.01{\rm cm^{-3}}$ and $n_0=0.001{\rm cm^{-3}}$, respectively. The corresponding values of $\epsilon_{\rm B}$ are $3\times10^{-4}$, $1\times10^{-3}$ and $4\times10^{-3}$, respectively for three lines. Other parameters are $E_{\rm II, iso}=2.2\times10^{53}$erg,  $p=2.4$, $\epsilon_{\rm e}=0.1$, and $\xi_{\rm e}=0.1$. 
    }
    \label{fig:spectra0}
\end{figure}
As $k_{\rm e}\equiv\epsilon_{\rm e}/\xi_{\rm e}$, a large $\epsilon_{\rm e}$ or small $\xi_{\rm e}$ will cause a high value of GeV-keV ratio.
The value of $k_e$ can be constrained by the spectral energy distribution  at $T_0+2.5{\rm d}$, as shown in Figure \ref{fig:spectra0}. The radio to X-ray afterglows are produced by the synchrotron emission of accelerated electrons. 
From  Figure \ref{fig:spectra0}, we can see that the radio to X-ray emissions do not follow a single power law, indicating the presence of a spectral break between  radio and X-ray frequencies.
A plausible explanation is that the break  $\nu_{\rm m}$  lies at  $\nu_{\rm m}\geq3\times10^{12}{\rm Hz}$ (The possibility that the break is the self-absorption frequency $\nu_{a}$  is disfavored, see the Appendix \ref{App:absorp}). The inequality arises from that the radio emission could include  contributions from both forward shock emission and reverse shock emission.  
As the break frequency $\nu_{\rm m}$  is given by
\begin{equation}
\nu_{\rm m}= 2.3{\rm GHz} E^{1/2}_{\rm\Rom{2},iso,54}k^2_{\rm e,-1.5}\epsilon ^{1/2}_{\rm B,-4}t^{-3/2}_{5.3},
\end{equation} 
we obtain
\begin{equation}
    \label{minke}
    k_{\rm e}\geq 1.1 E^{-1/4}_{\rm\Rom{2},iso,54}\epsilon^{-1/4}_{\rm B,-4}.
\end{equation}
With this lower limit of $k_{\rm e}$, we then obtain an upper limit on the circum-burst density by using Equation (\ref{ratio}), 
\begin{equation}
    \label{lowden}
    n_0 \leq 0.007{\rm cm^{-3}} E^{-\frac{3-p}{7-p}}_{\rm\Rom{2},iso,54} \epsilon^{\frac{2(p-1)}{7-p}}_{\rm B,-4}.
\end{equation}

\begin{figure*}
    \centering
    \includegraphics[scale=0.45]{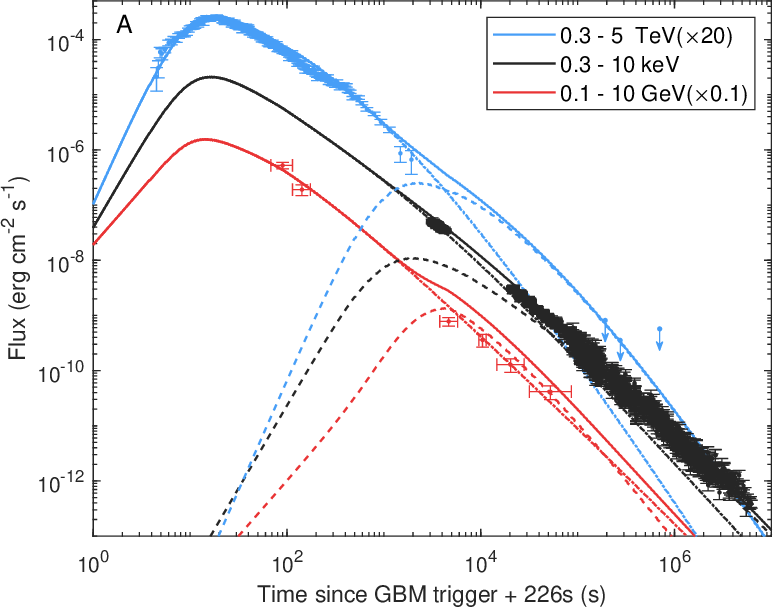}
    \includegraphics[scale=0.45]{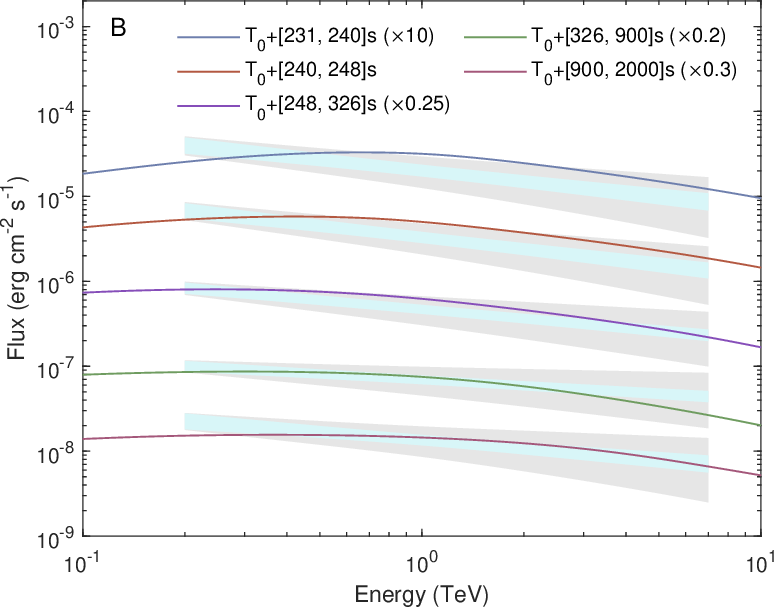}\\
    \includegraphics[scale=0.45]{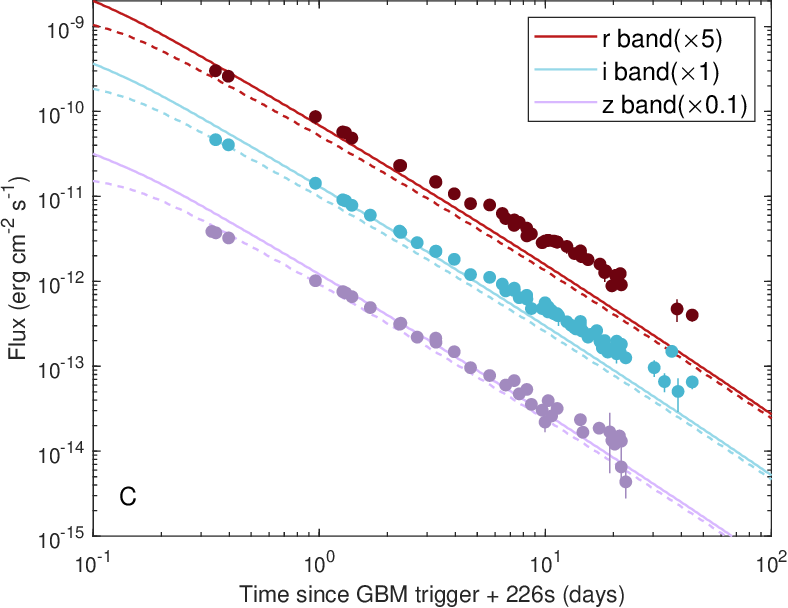}
    \includegraphics[scale=0.45]{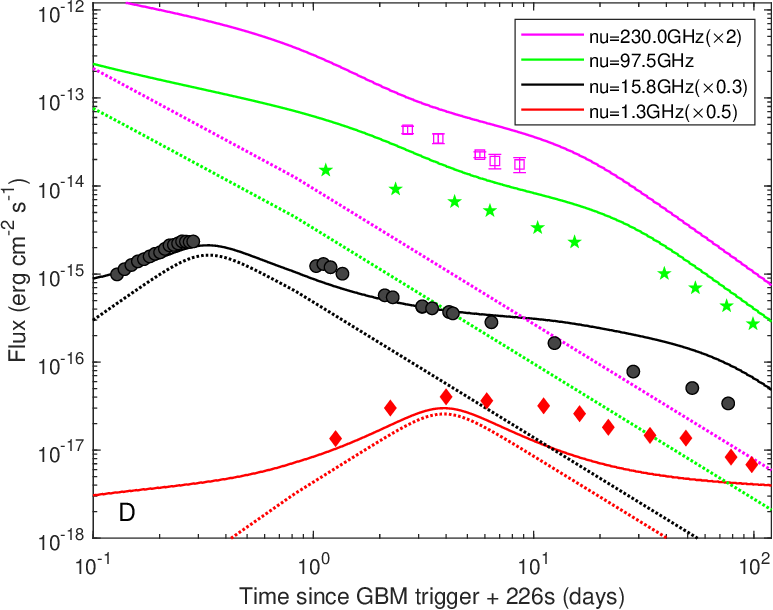}
    \caption{
    Modeling of the multi-wavelength data of GRB 221009A with  the standard afterglow theory. 
    Panel A displays the light curves in the energy band of TeV (0.3-5\,TeV), keV (0.3-10\,keV) and GeV (0.1-10\,GeV). The upper limits after $10^{5}$s show the H.E.S.S. data in the energy range 0.65-10 TeV.   The solid lines represent the sum of forward shock emission from the narrow core (dot-dashed lines) and the wide wing (dashed lines). 
    Panel B displays the spectra between 0.2-7 TeV measured by LHAASO in different time intervals \citep{LHAASO2023Sci...380.1390L}. The light blue band and grey band denote the systematic uncertainties and EBL-related uncertainties, respectively. The solid line represents the sum of forward shock SSC emission of the narrow core and wide wing. 
    Panel C displays the optical light curves in r band (red), i band (blue) and z band (purple). All optical data points are corrected only by the galactic extinction \citep{Schlafly2011ApJ...737..103S}. The solid lines represent the sum of forward shock emission from the narrow core and the wide wing (dashed lines). For brevity, the lines for the emission from the narrow core are not shown.
    Panel D shows the radio light curves at 230GHz (pink square), 97.5GHz (green star), 15.8GHz (black circle) and 1.5GHz (red diamond). The solid lines represent the sum of forward shock emission from the narrow core, the forward shock emission from the wide wing, and the reverse shock emission of the wide wing.  The dotted line represents the reverse shock emission of the wide wing. 
    }
    \label{fig:xie}
\end{figure*}

Such a low density  is dramatically different from the constraint we derived from the early TeV afterglow $n_{0}\geq0.1{\rm cm^{-3}}E^{-0.36}_{\rm\Rom{1},iso,55}\epsilon^{-0.5}_{\rm B,-4}$. It is hard  to reconcile the discrepancy by increasing $\epsilon_{\rm B}$, because    $h\nu_{\rm c}\propto E^{-1/2}_{\rm\Rom{1}, iso}\epsilon^{-3/2}_{\rm B}n^{-1}_{0}$ is more sensitive to $\epsilon_{\rm B}$ and the requirement $h\nu_{\rm c}\geq10{\rm keV}$ at $T_0+4000$s leads to
\begin{equation}
    \label{Xray}
    n_0 \leq 0.25{\rm cm^{-3}} E^{-1/2}_{\rm\Rom{1},iso,55} \epsilon^{-3/2}_{\rm B,-4}.
\end{equation}

The H.E.S.S. upper limit at 2.5 day also implies a low circum-burst density. In TeV band, the flux is suppressed by the KN effect and internal $\gamma\gamma$ absorption, so it can only be studied numerically. We model the SED of afterglows at 2.5 days in Figure \ref{fig:spectra0} with the synchrotron plus SSC emission, and find that only when $n_0<0.01{\rm cm^{-3}}$ the model can fit the data.

\subsection{A Stratified Density Profile: Transition  from a Constant-density Medium to a Wind Medium}
Therefore, we consider a  stratified density profile given by 
\begin{equation}
\label{density1}
n(r)= \left\{
\begin{array}{ll}
     n_0,  & r<r_{\rm c} \\
     A r^{-2}. & r\ge r_{\rm c}
\end{array}
\right.
\end{equation} 
The transition radius from constant-density region to the wind region occurs at $r_{\rm c}=\sqrt{A/n_0}=5.5\times10^{17}{\rm cm}A^{1/2}_{\star,-1}n^{-1/2}_{0,-1}$, corresponding to a transition time $t_{\rm c}=210E^{-1}_{\rm\Rom{1}, iso,55}A^{2}_{\star,-1}n^{-1}_{0,-1}$s. We assume the transition time is later than the jet break time $T^*$+670s, so the wind parameter should satisfy
\begin{equation}
    \label{highden2}
    A_{\star}\geq0.17E^{1/2}_{\rm\Rom{1},iso,55}n^{1/2}_{0,-1}.
\end{equation}

The standard wind-like profile ($k=2$) is based on the assumption of a constant mass loss rate $\dot{M}$ and wind velocity $v_{\rm w}$.
However, the dynamics of the stellar wind right before the death of the massive star is highly uncertain. Strong deviations from a density slope of $k=2$ may be expected  in the region very close to the GRB progenitor,  due to the short evolutionary timescales after core helium exhaustion and the effects of near-critical rotation, which can significantly alter wind properties. 
For example, if $\dot{M} \propto t^a$  and  and $v_w\propto t^b$, we have $\rho\propto r^{-2+(a-b)/(1+b)}$. A special mass loss rate and wind velocity ($a=3b+2$)  during the last few hundred years of the star’s life can lead to $k=0$ in the region  close to the GRB progenitor \citep{Yoon2006A&A...460..199Y,Colle2012ApJ...751...57D}. In addition, if the GRB occurs in massive stellar clusters,  the circum-burst medium is much more complicated due to the colliding wind effect \citep{Mimica2011MNRAS.418..583M}. 
In this case, one would expect an enhanced density due to the shocked colliding winds and a freely expanding wind at a larger distance.

\section{Model Fits of the Multi-wavelength Afterglow Data}
We consider an on-axis two-component jet model with the angular distribution described in Equation (\ref{eq:epsilon-structure}) expanding into a stratified circum-burst medium (Equation (\ref{density1})). The model  uses shock dynamics given in the Appendix \ref{app:num}. 
We assume that a fraction $\xi_{\rm e}$ of  shock-heated electrons are accelerated into a power law distribution with a spectral index of $p$: $dN_{\rm e}/d\gamma^{\prime}_{\rm e}\propto {\gamma^{\prime}_{\rm e}}^{-p}$, where  $\gamma^{\prime}_{\rm e}$ is the electron Lorentz factor. The complete KN cross section for the inverse Compton scattering has been considered and the internal $\gamma\gamma$ absorption within the emitting region has been taken into account (see the Appendix \ref{app:num}).

For the narrow core, we use the isotropic energy $E_{\rm\Rom{1},iso}=4\pi\epsilon_{\rm\Rom{1}}$ to solve the shock dynamics.  
For the wing component, since the energy has an angular distribution, it is reasonable to use the average isotropic energy in the solid angle between $\theta_{\rm j}$ and $\theta$ in the calculation, which is given by
\begin{equation}
    \bar{E}_{\rm\Rom{2},iso}(\theta)=\frac{4\pi\int^{\theta}_{\theta_{\rm j}}\epsilon_{\rm\Rom{2}}f(\theta^{\prime})\sin\theta^{\prime}d\theta^{\prime}}{\int^{\theta}_{\theta_{\rm j}}\sin\theta^{\prime}d\theta^{\prime}}.
\end{equation}
$\bar{E}_{\rm\Rom{2},iso}(\theta)$ starts to increase from $\theta_{\rm j}$ and decreases as a power-law $\bar{E}_{\rm\Rom{2},iso}\propto\theta^{-a}$ when $\theta\gg\theta_{\rm j}$.

\subsection{The Optical to TeV Afterglows}
The parameters of the angular profile of the wide wing can be estimated from the X-ray afterglow light curve analytically.
For a wing with angular profile $dE/d\Omega\propto \theta^{-a}$, the light curve of the synchrotron emission from the forward shock is given by \citep{Beniamini2022MNRAS.515..555B,Zhang&Wang2023}
\begin{equation}
\label{forward}
F_{\nu}\propto \left\{
\begin{array}{ll}
t^{-\frac{a}{3(4-a)}}, \,\,\,\, &\nu<\nu_m \\

t^{-\frac{2(3p-1)-a(p-1)}{2(4-a)}} , \,\,\,\, &\nu_m<\nu<\nu_c  \\

t^{-\frac{2(3p-2)-a(p-2)}{2(4-a)}} . \,\,\,\, & \nu>\nu_c
\end{array}
\right.
\end{equation} 
Before 0.8 d, the power-law slope of the X-ray afterglow is $\alpha=-1.52$, suggesting an angular profile index of $a=0.2$. After 0.8 d, the slope becomes steeper with $\alpha=-1.66$, leading to $a=0.7$. 

The bulk Lorentz factor is $\Gamma=23E^{1/4}_{\rm\Rom{2},iso,54}A^{-1/4}_{\star,-1}$ at 0.8 d, suggesting that the transition angle from a shallow angular profile
to a steeper angular profile is $\theta_{\rm c,w}\sim\Gamma^{-1}\approx2.5^{\circ}E^{-1/4}_{\rm\Rom{2},iso,54}A^{1/4}_{\star,-1}$. 

We model the multi-wavelength afterglow data of GRB 221009A with the two-component jet model, which is shown in Figure \ref{fig:xie}. We first assume that there is no lateral expansion for both jet components.  We find that this model can explain the TeV, X-ray and optical afterglows, with  the model parameter values given in Table \ref{notation}. 

The early TeV afterglow emission originates from the SSC emission of the narrow jet component. At times later than $10^4{\rm s}$, the SSC emission from the wing component becomes dominant, and its flux is  consistent with the  upper limits imposed by H.E.S.S. observations. The X-ray afterglow at times later than $10^4{\rm s}$ is produced by the wing component through the synchrotron emission. The optical afterglow is also produced by the synchrotron emission of the wing component. The model flux is insufficient to explain the optical data after tens of days, which could be attributed to extra contribution from a supernova \citep{Fulton2023ApJ...946L..22F,Levan2023ApJ...946L..28L,Srinivasaragavan2023ApJ...949L..39S,Blanchard2023arXiv230814197B}.    

\begin{figure*}
    \centering
    \includegraphics[scale=0.45]{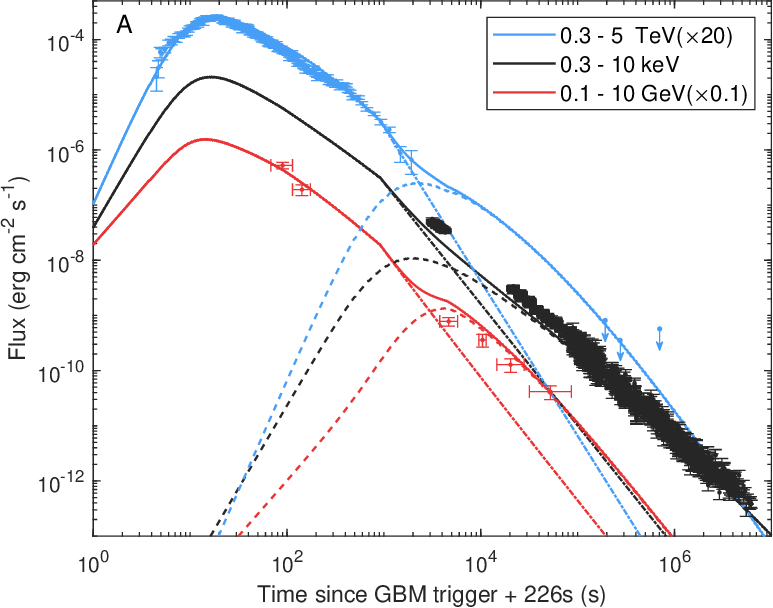}
    \includegraphics[scale=0.45]{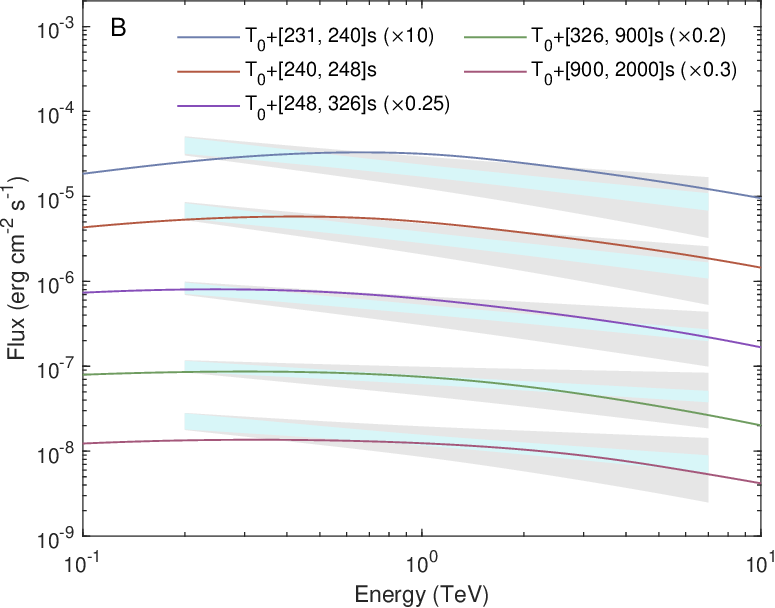}\\
    \includegraphics[scale=0.45]{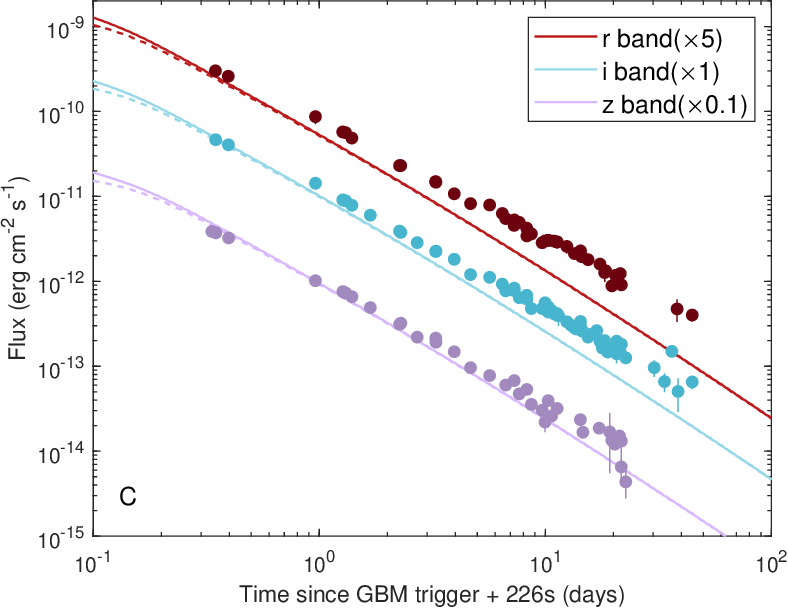}
    \includegraphics[scale=0.45]{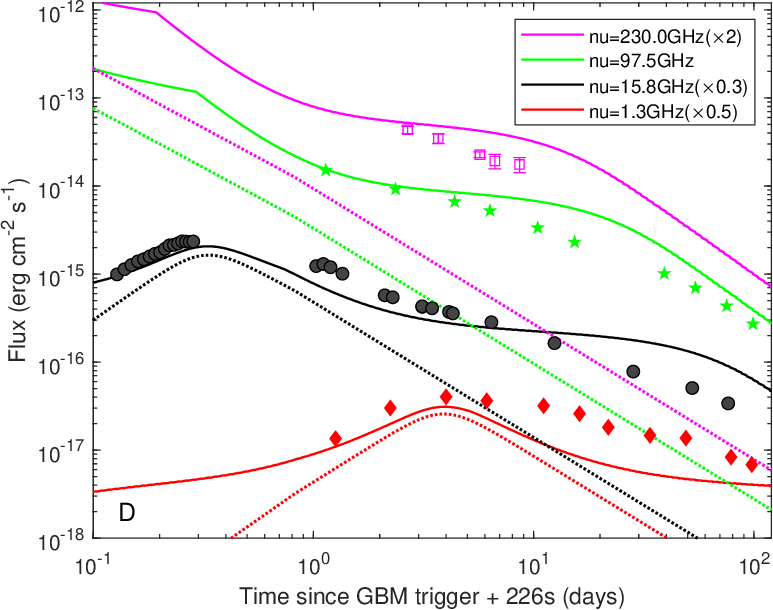}
    \caption{Same as Figure \ref{fig:xie}, but assuming that the narrow core has lateral expansion when the bulk Lorentz factor drops to  $\Gamma_{\rm \Rom{1}}\leq \theta^{-1}_{\rm j}$.  
    }
    \label{fig:xieLat}
\end{figure*}

The isotropic energy of the narrow core is $4\pi \epsilon_{\rm I}=9\times 10^{54}{\rm erg}$, while the isotropic energy of the wing at $\theta=\theta_{\rm c,w}$ is $4\pi \epsilon_{\rm\Rom{2}}/\sqrt{2}=2.8\times 10^{53}{\rm erg}$.  The indices of the angular profile of the wing,  $a_1=0$ and $a_2=0.8$,  are consistent with the above analytical estimate. The acceleration fraction $\xi_{\rm e}=0.15$ for the wide wing is an order of magnitude higher than that in \cite{Connor2023SciA....9I1405O} and \cite{Gill2023MNRAS.524L..78G}, which  avoids to overproduce the TeV flux measured by H.E.S.S. at 2.5 days. 

The theoretical flux at GeV band slightly exceeds the observational data at $\sim 10^4{\rm s}$, during which both the narrow jet and wide wing contribute to the  flux. 
Semi-analytical models and numerical simulations have shown that lateral expansion may be  important for narrow jets \citep{Granot2012MNRAS.421..570G,Lu2020arXiv200510313L}, so our calculation may overestimate the emission flux from the narrow core component after its jet break time if the later expansion occurs in structured jets \citep{Gottlieb2021MNRAS.500.3511G}. The model considering the lateral expansion for the narrow core is shown in Figure \ref{fig:xieLat}, which now agrees better with the GeV data.

\subsection{The Radio Afterglow}
The radio afterglow of GRB 221009A may be produced by both the forward shock emission and reverse shock emission. Since the narrow jet is likely to be Poynting-flux-dominated,  we do not consider the reverse shock emission from this component.  The reverse shock emission in our model comes from the wide wing.

The early radio data around 0.2 d exhibit a spectral shape $F_{\nu}\propto\nu^{5/2}$, indicating that the observed frequency ($\sim15$GHz) is between the self-absorption frequency and the minimum frequency, i.e., $\nu_{\rm m}<\nu<\nu_{\rm a}$. This feature is inconsistent with the forward shock emission, which has a much higher $\nu_{\rm m}$. Therefore, the early radio is probably dominated by  the reverse shock.

Since the deceleration time of the wide jet is $\sim$3000s, the wide jet can be regarded as a thin shell. After the deceleration, the Lorentz factor of reverse shock follows the relation $\Gamma_{3}\propto R^{-g}$. For the wind environment, we  take $g=1$, and the rising slope of reverse shock emission in the spectral regime of $\nu_{\rm m,rs}<\nu<\nu_{\rm a,rs}$ is $\alpha=\frac{5(8+5g)}{14(1+2g)}=1.55$ \citep{Zou2005MNRAS.363...93Z}, which is close to the observed rising slope $\alpha=1.4$ \citep{Bright2023NatAs...7..986B}.

However, the observed decay slope  of the radio afterglow, $\alpha=-0.8$, is much shallower than the prediction by the reverse shock in a top-hat jet, which is $\alpha\approx-2$. \cite{Zhang&Wang2023} studied the reverse shock evolution in a structured jet and found the decay slope can be shallower when the initial Lorentz factor of the jet has an angular structure, i.e., $\Gamma_0\propto\theta^{\rm -k_\Gamma}$ (see Appendix \ref{App:Rev}).   Considering this effect, the decaying slope of the reverse shock emission is $\alpha=-1.35$ for $a=0$ and $\alpha=-1.54$ for $a=0.8$, as discussed in Appendix \ref{App:Rev}.

From Figure \ref{fig:xie} and \ref{fig:xieLat}, we can see that the model does not explain the radio afterglow data satisfactorily. The forward shock emission exceeds the high-frequency radio data by a factor of about 2 in the case assuming no  lateral expansion for the narrow jet and by a factor of 1.5 assuming lateral expansion.   

Indeed, it is found that the light curves of radio afterglows of many GRBs can not be explained satisfactorily by the standard afterglow theory \citep[e.g.][]{Levine2023MNRAS.519.4670L}. This could indicate our lack of knowledge about the radio afterglow. Below we discuss some non-standard scenarios  that could possibly resolve the discrepancy in the radio afterglows of GRB 221009A.

\section{Non-standard Afterglow Model}
In the standard model of afterglows, the micro-physical parameters $\epsilon_{\rm e},\epsilon_{\rm B},\xi_{\rm e}$ and $p$ are usually assumed to be a constant in the whole afterglow. The assumption simplifies the modeling and can explain afterglows of many GRBs. However, for those GRBs which are thoroughly investigated (e.g. GRB 130427A $\&$ GRB 170817A), time-evolving microphysics parameters are proposed to explain some unusual behaviour of afterglows \citep[e.g.][]{Maselli2014Sci...343...48M,Takahashi2022MNRAS.517.5541T}.

For GRB 221009A, the radio observations after 10 days impose a  stringent limit on $\nu_{\rm m}$ and $k_{\rm e}$. 
At $T_0+54$ days, the spectrum between X-ray and radio is not a single power-law, so $\nu_{\rm m}$ is required to be higher than $3\times10^{11}{\rm Hz}$, which gives
\begin{equation}
    \label{minke2}
    k_{\rm e}\geq 1.7 E^{-1/4}_{\rm\Rom{2},iso,54}\epsilon^{-1/4}_{\rm B,-3}.
\end{equation}
On the other hand, we obtain an upper limit of $k_{\rm e}\lesssim0.4$ at 2.5 d from the SED modeling for $A_{\star}=0.17$ and $E_{\rm\Rom{2},iso}=10^{54}{\rm erg}$. The discrepancy of $k_{\rm e}$ at the two times can not be solved  by increasing $\epsilon_{\rm B}$ in Equation (\ref{minke2}), because the X-ray spectra at $T_{0}+26000$s requires $h\nu_{\rm c}\geq10{\rm keV}$, resulting in $\epsilon_{\rm B}\leq1.2\times10^{-3}E^{1/4}_{\rm \Rom{2},iso,54}A^{-4/3}_{\star,-1}$.  
Furthermore, the isotropic equivalent energy decreases with the time as $E_{\rm \Rom{2},iso}\propto t^{-\frac{a}{4-a}}$ for a structured jet (see the Appendix \ref{App:Rev}), then a higher value of $k_{\rm e}$ is needed at 54 days. Thus, we speculate that $k_{\rm e}$ may increase with time in GRB 221009A. As $k_{\rm e}=\epsilon_{\rm e}/\xi_{\rm e}$, the increase of $k_{\rm e}$  can be due to a decrease of $\xi_{\rm e}$ or a increase of $\epsilon_{\rm e}$.

\subsection{Decreasing Acceleration Fraction $\xi_{\rm e}$ }
The fraction of particles being accelerated in afterglow shocks depends on the conditions of the shock, including the bulk Lorentz factor, the magnetization parameter and the direction of the magnetic field \citep{Sironi2015SSRv..191..519S}. Particle in cell (PIC) simulations find that the fraction of non-thermal electrons is about $\xi_{\rm e}\sim 1-10\%$  \citep{Spitkovsky2008ApJ...682L...5S,Sironi2013ApJ...771...54S}. These simulations are conducted with the bulk Lorentz factor $\Gamma=10-20$, which is different from the condition at early time when $\Gamma_0> 100$. On the other hand, the magnetization degree can influence the direction of the magnetic field, possibly changing the acceleration fraction as well \citep{Sironi2013ApJ...771...54S}.
Thus the acceleration fraction $\xi_{\rm e}$ could be time-varying. 

The high-frequency radio data ($\nu>90{\rm GHz}$) lies above the power-law extrapolation of $F_{\nu}\propto\nu^{-0.2}$ from 1-20 GHz (see Figure 5 in \cite{Laskar2023ApJ...946L..23L}), implying that the forward shock may contribute to these bands. A decreasing $\xi_{\rm e}$ is helpful to avoid the overshoot in the radio flux at late times, because the synchrotron emission from the forward shock,  $F_{\nu}\propto\epsilon^{-2/3}_{\rm e}\xi^{5/3}_{\rm e}$, is  sensitive to $\xi_{\rm e}$ in the spectral regime of $\nu<\nu_{\rm m}$. For the X-ray and optical band, the evolution of $\xi_{\rm e}$ affects the light curve only slightly, as $F_{\nu}\propto\epsilon^{p-1}_{\rm e}\xi^{2-p}_{\rm e}$.

\begin{figure*}
\centering
    \includegraphics[scale=0.45]{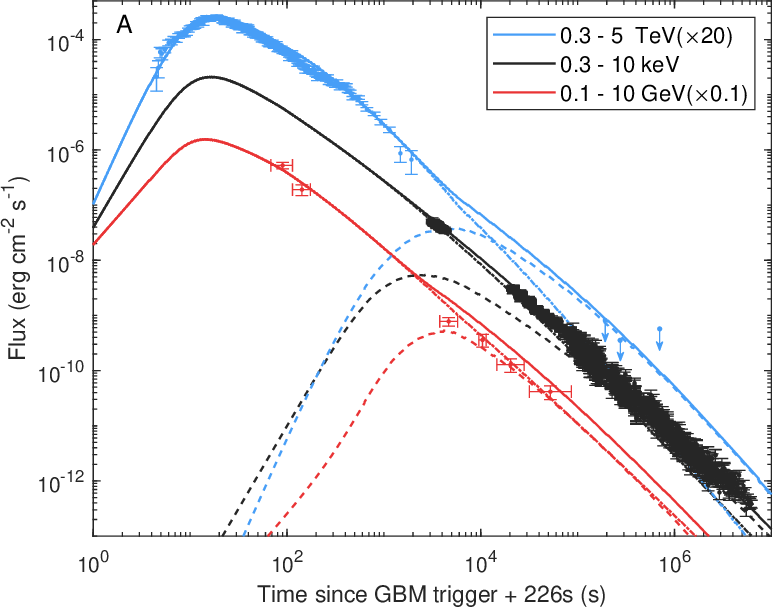}
    \includegraphics[scale=0.45]{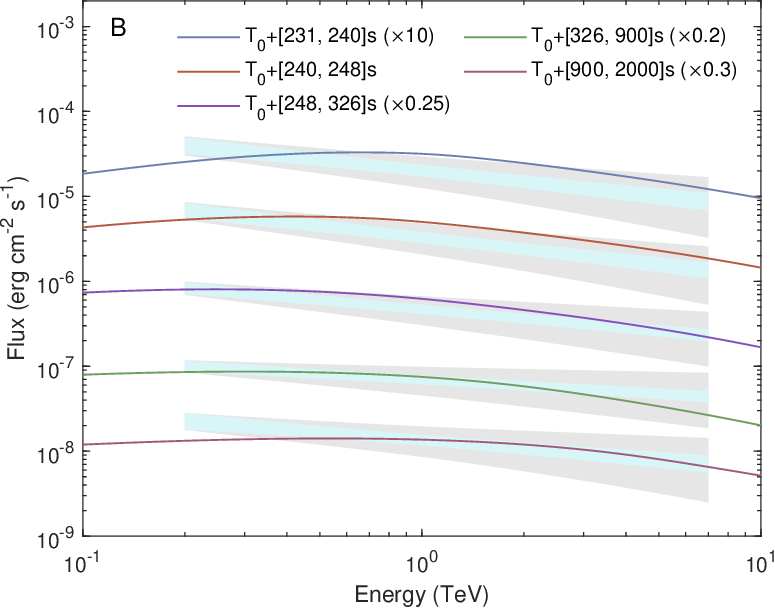}\\
    \includegraphics[scale=0.45]{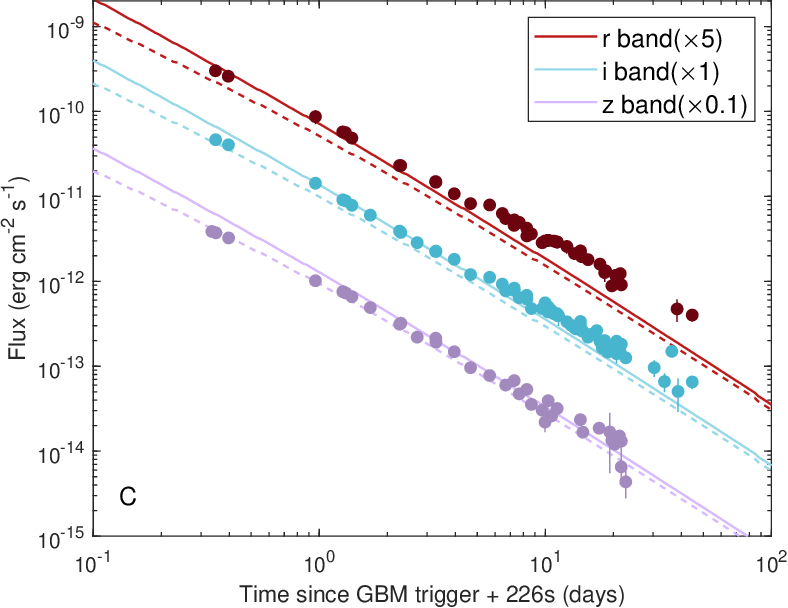}
    \includegraphics[scale=0.45]{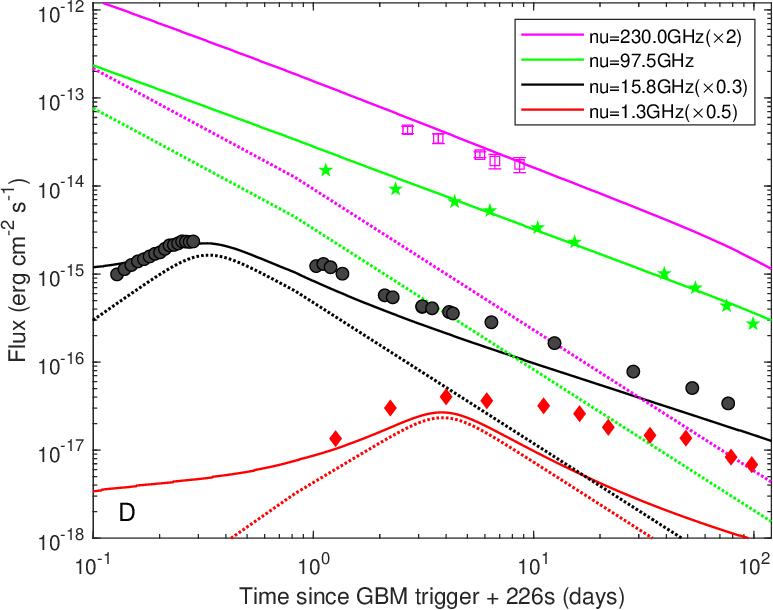}
    \caption{Same as Figure \ref{fig:xie}, but assuming that the acceleration fraction $\xi_{\rm e}$ in two components are decreasing with time. 
    }
    \label{fig:wind}
\end{figure*}

We assume the fraction of accelerated electrons decrease  with time as $\xi_{\rm e}\propto t^{-\alpha_{\xi}}$. In wind environment, the light curve of the structured jet is 
\begin{equation}
F_{\nu}\propto \left\{
\begin{array}{ll}
t^{-\frac{a}{3(4-a)}-\frac{5}{3}\alpha_{\xi}}, \,\,\,\, &\nu<\nu_m \\

t^{-\frac{2(3p-1)-a(p-1)}{2(4-a)}+\alpha_{\xi}(p-2)} , \,\,\,\, &\nu_m<\nu<\nu_c  \\

t^{-\frac{2(3p-2)-a(p-2)}{2(4-a)}+\alpha_{\xi}(p-2)} . \,\,\,\, & \nu>\nu_c
\end{array}
\right.
\end{equation} 
The spectral index of X-ray afterglow measured by Swift XRT at later epochs is $\beta\approx-0.78$, suggesting a power-law index of $p\simeq 2.5$. After 0.8 d, the decay slopes of X-ray and high-frequency radio afterglows are $\alpha=-1.66$ and $\alpha=-0.78$ respectively. To fit the multi-wavelength data simultaneously, we need $a=0.9$ and $\alpha_{\xi}=0.4$. Because the bulk Lorentz factor of the afterglow shock decreases with time as $\Gamma\propto t^{-1/(4-a)}$, we then derive $\xi_{\rm e}\propto \Gamma^{-1.24}$.
Before 0.8 d, the decay slope of the X-ray afterglow is $\alpha=-1.52$, suggesting an angular profile index of $a=0.2$. 
We assume $\xi_{\rm e}=1$ when the Lorentz factor is greater than 60.

\begin{table*}
 \centering
 \caption{Parameter values used in the modeling of the multi-wavelength afterglow data. The subscript ${\rm\Rom{1}}$ and ${\rm\Rom{2}}$ represent the parameters for the narrow core and wide wing, respectively. Forward shock1 (FS1) denotes the standard forward shock model shown in Figure \ref{fig:xie}, while Forward shock2 (FS2) denotes the non-standard forward shock model with a time-varying $\xi_{\rm e}\propto\Gamma^{-\alpha_{\Gamma}}$, which was shown in Figure \ref{fig:wind}. We use $\alpha_{\rm\Rom{1},\Gamma}=1.35$ and $\alpha_{\rm\Rom{2},\Gamma}=1.25$ in the non-standard case. 
 The parameter values for the density profile are $n_0=0.2{\rm cm^{-3}}$ and $A_{\star}=0.17$.}
 \begin{tabular}{cccccccccccccccccc}
 \hline
 \hline
  &
 $4\pi\epsilon_{\rm\Rom{1}}({\rm erg})$ & $\theta_{\rm j}$ & $\Gamma_{\rm\Rom{1},0}$ & $\epsilon_{\rm\Rom{1},e}$ & $\epsilon_{\rm\Rom{1},B,-3}$ & $\xi_{\rm\Rom{1},e}$ & $p_{\rm\Rom{1}}$ & $4\pi\epsilon_{\rm\Rom{2}}({\rm erg})$ &   $\theta_{\rm c,w}$ & $\Gamma_{\rm\Rom{2},0}$ &  $a_1$ & $a_2$ & $\epsilon_{\rm\Rom{2},e}$ &  $\epsilon_{\rm\Rom{2},B,-3}$ &
  $\xi_{\rm\Rom{2},e}$ &$p_{\rm\Rom{2}}$   \\
 \hline

 FS1 &
 $9\times10^{54}$ & $0.6^{\circ}$ & 560 & 0.04 & 1 & 1 & 2.2 & $4\times10^{53}$ & $3^{\circ}$ & 60 & 0 & 0.8 & 0.06  & 2 & 0.15 & 2.4 \\

 FS2 &
 $9\times10^{54}$ & $0.6^{\circ}$ & 560 & 0.04 & 1 & --- & 2.2 & $4\times10^{53}$ & $3^{\circ}$ & 60 & 0 & 1.0 & 0.06  & 2 & --- & 2.5 \\

 RS &
 --- & --- & --- & --- & --- & --- & --- & 
 $4\times10^{53}$ & $3^{\circ}$ & 60 & 0 & 0.8/1.0\footnote{This parameter follows the forward shock.} & 0.04  & 0.5 & 0.01 & 2.2 \\
 
 \hline
 \end{tabular}
 \label{notation}
\end{table*}

We model the multi-wavelength afterglows with a time-varying $\xi_{\rm e}$, which is  shown in Figure \ref{fig:wind}. The fitting of radio data is improved. For low-frequency radio afterglows, the model flux is still below the data at later times. The excess  could arise from some extra low-velocity ejecta components in this GRB, which may produce main radio emissions. 

Since $\epsilon_{\rm e}$ and $\xi_{\rm e}$ are coupled in the afterglow modeling, an alternative model is that $\epsilon_{\rm e}$ is increasing with time. The main difference lies in that   the X-ray and optical flux depends on $\epsilon_{\rm e}$ as $F_{\nu}\propto\epsilon^{p-1}_{\rm e}$. To fit the X-ray and optical light curves, we would need a steeper angular profile for the jet structure with $a\approx2$.

\subsection{Decaying Magnetic Field}
In the standard afterglow shock model, we assume a homogeneous magnetic field in the downstream of the  shock. Nonetheless, the realistic magnetic field may have a spatial distribution behind the shock.
PIC simulations and theoretical analyses of relativistic collisionless shocks both suggest that the coherent length of magnetic fields  is much smaller than the shock size of GRB afterglows \citep{Chang2008ApJ...674..378C,Lemoine2015JPlPh..81a4501L,Sironi2015SSRv..191..519S}.

Since the cooling time of radio-emitting electrons is much longer than the shock dynamic time,
these electrons may radiate most of their energy at the back of the blast wave, where the magnetic field has decayed to a low value \citep{Lemoine2013MNRAS.428..845L,Wang2013ApJ...771L..33W}. The standard model assuming a constant $\epsilon_{\rm B}$ may overestimate the radio flux at later epochs. Taking a realistic magnetic field might reduce the radio flux and solve the discrepancy.

\section{Conclusions and Discussions}
GRB 221009A,  as the BOAT event, provides rich multi-wavelength afterglow data spanning from GHz to TeV energies. We find that the late-time multi-wavelength observations, including radio, X-ray, GeV and TeV, combine to constrain the density of the circum-burst medium to be lower than $n_0<0.01{\rm cm^{-3}}$, while the LHAASO observation at early time requires a constant-density medium with  $n_{0}\geq0.1{\rm cm^{-3}}E^{-\frac{9-p}{21-p}}_{\rm \Rom{1},iso,55}\epsilon^{-\frac{2(7-p)}{21-p}}_{\rm B,-4}$ at small radii. Therefore,  we propose a  stratified density profile that incorporates a constant-density medium at small radii and a wind-like medium at large radii to explain the afterglows of GRB 221009A.

Motivated by the multi-wavelength data, we employ a two-component jet model, comprising a uniform narrow jet core and a structured wing. This model can explain the afterglows  from optical to TeV bands, although the  flux at high-frequency radio bands exceeds the data by a factor of two after the second day (see Figure \ref{fig:xie} and \ref{fig:xieLat}). We find that the discrepancy could be resolved by invoking  time-varying  micro-physical parameters of afterglow shocks (see Figure \ref{fig:wind}).

The model flux is lower than the observed flux at low-frequency radio bands (e.g., $\nu=1.3$GHz) after $\sim$2 days (see Figure \ref{fig:wind}). The excess could be due to some extra electron component, possibly resulting from  low-velocity ejecta components or related to other acceleration mechanisms. The hard spectra of low-frequency radio data $\beta\approx-0.2$ is unusual in GRB afterglows, suggesting a hard injection spectra $p=1.4$ for electrons, which is in contradiction to standard diffusive shock acceleration theory. Such a hard spectrum could result from some other acceleration mechanism, such as shear acceleration \citep{Liu2017ApJ...842...39L, Rieger22}.

Compared to previous single structured jet models \citep{Connor2023SciA....9I1405O,Gill2023MNRAS.524L..78G},
our  two-component structured jet model explains TeV and lower-energy afterglows simultaneously. Previous single structured jet models did not take into account the SSC process because the TeV data were not available.  We calculate the SSC emission and include the TeV data in the modeling, providing a more self-consistent explanation for the multi-wavelength afterglows.

In our model, the beaming-corrected kinetic energy in the narrow core is  $E_{\rm\Rom{1},b}=E_{\rm\Rom{1},iso}(1-\cos\theta_{\rm j})=5\times10^{50}{\rm erg}$ \citep{LHAASO2023Sci...380.1390L}. The kinetic energy in the structured wing is  $E_{\rm\Rom{2},b}=2\int^{\theta}_{\theta_{\rm j}}\frac{dE}{d\Omega}d\Omega$, where $\theta$ is the angle corresponding to the last observing time $t$ before seeing the maximum angle $\Theta$. The beaming-corrected energy increase as $E_{\rm\Rom{2},b}\propto\theta^2\propto t^{\frac{2-a}{4-a}}$\citep{Beniamini2022MNRAS.515..555B}. 
We find the beaming-corrected energy of the structured wing is $E_{\rm\Rom{2},b}=5.7\times10^{51}(t/100{\rm days})^{3/8}{\rm erg}$ for $a_2=0.8$ and  $E_{\rm\Rom{2},b}=4.9\times10^{51}(t/100{\rm days})^{1/3}{\rm erg}$ for $a_2=1$. These values are significantly lower than the energy budget required by  single structured jet models, which are $8\times10^{52}(t/80{\rm days})^{0.372}{\rm erg}$ in the model of \cite{Connor2023SciA....9I1405O} and $4\times10^{52}(t/100{\rm days})^{0.375}{\rm erg}$ in the model of \cite{Gill2023MNRAS.524L..78G}, respectively.

\begin{acknowledgments}
The authors thank Katsuaki Asano, Hai-Ming Zhang, Liang-Duan Liu and Yu-Jia Wei for useful discussions. This work is supported by  the National Natural Science Foundation of China (grant numbers 12333006, 12121003, U2031105). 
\end{acknowledgments}

\setcounter{figure}{0}
\renewcommand{\thefigure}{A\arabic{figure}}

\appendix

\section{Methods}
\subsection{Analytical Methods}
\label{app:ana}
The analytical solution of the bulk Lorentz factor $\Gamma$ and the radius $R$ of relativistic blast wave in an arbitrary power-law density profile $n=Ar^{-k}$ are give by \citep{Granot2002ApJ...568..820G}

\begin{equation}
\label{ana1}
    \Gamma  = 1.15^{2-k}{\left( {\frac{{\left( {17 - 4k} \right)E_{\rm iso}}}{{{4^{5 - k}}{{\left( {4 - k} \right)}^{3 - k}}\pi A{m_{\rm p}}{c^{5 - k}}{t^{3 - k}_{\rm z}}}}} \right)^{\frac{1}{{2\left( {4 - k} \right)}}}}, \quad\quad
    R = 1.3^{-k-1}{\left( {\frac{{\left( {17 - 4k} \right)\left( {4 - k} \right)E_{\rm iso}t_{\rm z}}}{{4\pi A{m_{\rm p}}c}}} \right)^{\frac{1}{{4 - k}}}},
\end{equation}
where $t_{\rm z}=t/(1+z)$ is the time corrected by redshift, 1.15 and 1.3 are numerical correction factors. Assuming the injection spectra of electrons is a single power law $dN_{\rm e}/d\gamma^{\prime}_{\rm e}\propto{\gamma^{\prime}_{\rm e}}^{-p}$. The prime marks the comoving frame of the shock. The minimum Lorentz factor and the cooling Lorentz factor are 
\begin{equation}
\label{ana2}
    \gamma^{\prime}_{\rm m}=\frac{\epsilon_{\rm e}}{\xi_{\rm e}}\frac{p-2}{p-1}\frac{m_{\rm p}}{m_{\rm e}}(\Gamma-1), \quad\quad 
    \gamma^{\prime}_{\rm c}=\frac{6\pi m_{\rm e} c}{\sigma_{\rm T}\Gamma B^{\prime 2}t_{\rm z}},
\end{equation}
where $\epsilon_{\rm e}$ is the equapartition factor of electrons, and $\xi_{\rm e}$ is the fraction of accelerated electrons. The magnetic field in the comoving frame is $B^{\prime}=\sqrt{8\pi\epsilon_{\rm B}nm_{\rm p}c^2(\Gamma-1)(\hat{\gamma}\Gamma+1)/(\hat{\gamma}-1)}$, where $\hat{\gamma}$ is the adiabatic index of the shock. We use the fitting formula for $\hat{\gamma}$ from \cite{Pe2012ApJ...752L...8P}, which is $(5-1.21937z+0.18203z^2-0.96583z^3+2.32513z^4-2.39332z^5+1.07136z^6)/3$, where $z=\zeta/(0.24+\zeta)$, $\zeta=(\frac{\Gamma\beta_{\rm sh}}{3})(\frac{\Gamma\beta_{\rm sh}+1.07(\Gamma\beta_{\rm sh})^2}{1+\Gamma\beta_{\rm sh}+1.07(\Gamma\beta_{\rm sh})^2})$, and
$\beta_{\rm sh}=\sqrt{1-\Gamma^{-2}}$ is the velocity of the shock.

In a constant medium($k=0$), the characteristic break frequencies in the synchrotron emission spectrum are
\begin{equation}
    \nu_{\rm m}=\frac{\Gamma{\gamma^{\prime}_{\rm m}}^2\nu^{\prime}_{\rm L}}{1+z}={2.4 \times {10}^{11}{\rm Hz} E_{\rm iso,55}^{1/2}k_{\rm e,-1}^2\epsilon _{\rm B,-4}^{1/2}t_5^{-3/2}}, \quad
    \nu_{\rm c}=\frac{\Gamma{\gamma^{\prime}_{\rm c}}^2\nu^{\prime}_{\rm L}}{1+z}= 6.7\times10^{17}{\rm Hz} 
    E_{\rm iso,55}^{-1/2}n_{0,-0.5}^{-1}\epsilon_{\rm B,-4}^{-3/2}t_{5}^{-1/2},
\end{equation}
where $\nu^{\prime}_{\rm L}=eB^{\prime}/2\pi m_{\rm e}c$ is the Lamour frequency of electrons. The characteristic break energies for SSC emission are
\begin{equation}
    h\nu^{\rm IC}_{\rm m}=2{\gamma^{\prime}_{\rm m}}^2h\nu_{\rm m}=65{\rm GeV}{E_{\rm iso,55}^{3/4}k_{\rm e,-1}^4\epsilon _{\rm B,-4}^{1/2}n_{0,-0.5}^{-1/4}t_{1.5}^{-9/4}}, \quad
    h\nu^{\rm IC}_{\rm c}=2{\gamma^{\prime}_{\rm c}}^2h\nu_{\rm c}=28{\rm PeV}{E_{\rm iso,55}^{-5/4}\epsilon_{\rm B,-4}^{-7/2}n_{0,-0.5}^{-9/4}t_{1.5}^{-1/4}}.
\end{equation}
If $h\nu^{\rm IC}_{\rm c}\gtrsim\Gamma\gamma^{\prime}_{\rm c}m_{\rm e}c^2$, the cooling break energy of SSC flux is $E^{\rm IC}_{\rm c,KN}=0.2\Gamma\gamma^{\prime}_{\rm c}m_{\rm e}c^2=0.75{\rm TeV}E^{-1/4}_{\rm iso,55}\epsilon^{-1}_{\rm B,-4}n^{-3/4}_{0,-0.5}t^{-1/4}_{1.5}$ due to the KN effect \citep{Nakar2009ApJ...703..675N}.

In the observer frame, the peak flux density of the synchrotron  and SSC emission are
\begin{equation}
    F_{\rm \nu,max}=(1+z)\frac{N_{\rm e}P_{\rm \nu}}{4\pi D^2_{\rm L}}=7{\rm Jy}E_{\rm iso,55}\xi_{\rm e,0}\epsilon^{1/2}_{\rm B,-4}n^{1/2}_{0,-0.5}, \quad
    F^{\rm IC}_{\rm m}=\tau_{\rm IC}F_{\rm \nu,max}=0.19{\rm\mu Jy}E^{5/4}_{\rm iso,55}\xi_{\rm e,0}\epsilon^{1/2}_{\rm B,-4}n^{5/4}_{0,-0.5}t_{1.5}^{1/4},
\end{equation}
where $P_{\rm \nu}=\sqrt{3}e^3\Gamma B^{\prime}/m_{\rm e}c^2$ is the spectral power of synchrotron, $N_{\rm e}=\int4\pi r^2\xi_{\rm e}n(r)dr$ is the number of accelerated electrons, $D_{\rm L}=716$Mpc is the luminosity distance, $\tau_{\rm IC}=n_0\sigma_{\rm T}R/3$ is the optical depth of Inverse Compton.


\subsection{Numerical Methods}
\label{app:num} 

The dynamical equations of the blast wave is described by \citep{Huang1999MNRAS.309..513H}
\begin{equation}
    \frac{d\Gamma}{dm_{\rm sw}}=-\frac{{{\Gamma ^2} - 1}}{{{M_0} + \left[ {f + 2\Gamma \left( {1 - f} \right)} \right]m_{\rm sw}}}, \quad\quad
    \frac{dR}{dt}=\frac{\beta_{\rm sh} c}{1-\beta_{\rm sh}},
\end{equation}
where $M_0=E_{\rm iso}/(\Gamma_0-1)c^2$ is the initial mass of the ejcta, $m_{\rm sw}=\int4\pi r^2\rho(r)dr$ is the swept-up mass and $f$ is the radiative efficiency. We employ a constant isotropic energy $E_{\rm iso}=4\pi\epsilon_{\rm\Rom{1}}$ for the narrow jet and the average isotropic energy $E_{\rm iso}=\bar{E}_{\rm\Rom{2},iso}(\theta)$ for the wide jet. 

We adopt $f=\epsilon_{\rm e}(t^{\prime -1}_{\rm syn}(\gamma^{\prime}_{\rm m})+t^{\prime -1}_{\rm IC}(\gamma^{\prime}_{\rm m}))/(t^{\prime -1}_{\rm syn}(\gamma^{\prime}_{\rm m})+t^{\prime -1}_{\rm IC}(\gamma^{\prime}_{\rm m})+t^{\prime -1}_{\rm ad})$ in our calculations, where $t^{\prime}_{\rm ad}=R/\Gamma\beta_{\rm sh}c$ is timescale of adiabatic cooling, and $t^{\prime}_{\rm syn}(\gamma^{\prime}_{\rm m})$ and $t^{\prime}_{\rm syn}(\gamma^{\prime}_{\rm m})$  are cooling timescale of synchrotron and inverse Compton at $\gamma^{\prime}_{\rm m}$, respectively. 

For electron spectra, we include the cooling of synchrotron and inverse Compton in the calculation $\gamma^{\prime}_{c}=6\pi m_{e}c/\sigma\Gamma B^{\prime 2}t_{\rm z}(1+f_{\rm KN}Y)$, where $Y\equiv P_{\rm SSC}/P_{\rm syn}$ is the ratio between SSC power and synchrotron power. The KN suppress factor is defined as $f_{\rm KN}\equiv P^{\rm KN}_{\rm SSC}/P^{\rm T}_{\rm SSC}$, where the $P^{\rm T}_{\rm SSC}=\frac{4}{3}\pi\sigma_{\rm T}c\gamma^2U^{\prime}_{\rm syn}$. The total energy loss rate of Compton scattering in KN regime is \citep{Blumenthal1970RvMP...42..237B}
\begin{equation}
    {P^{\rm KN}_{\rm IC}} = 12{\gamma ^2}{\sigma _{\rm T}}{c\int_{{0}}^\infty  {\varepsilon^{\prime} \frac{{dn^{\prime}}}{{d\varepsilon^{\prime} }}d\varepsilon^{\prime} } \int_0^1 {\frac{{qG\left( {q,{\Gamma _e}} \right)}}{{{{\left( {1 + {\Gamma _e}q} \right)}^3}}}dq} },
\end{equation}
where $\Gamma_e=4\gamma\varepsilon^{\prime}/m_{\rm e}c^2$, $G\left( {q,{\Gamma _e}} \right)$ is a function of KN cross section and $dn^{\prime}/d\varepsilon^{\prime}$ is the differential number density of synchrotron photons.
\begin{equation}
    G\left( {q,{\Gamma _e}} \right) = 2q\ln q + \left( {1 + 2q} \right)\left( {1 - q} \right) + \frac{{\Gamma _e^2{q^2}\left( {1 - q} \right)}}{{2\left( {1 + {\Gamma _e}q} \right)}}.
\end{equation}
The spectral power of synchrontron emission is 
\begin{equation}
    P^{\prime}_{\nu}(\nu^{\prime})=\frac{\sqrt{3}eB^{\prime}}{m_{\rm e}c^2}\int^{\infty}_{\gamma_{\rm m}} \left(\frac{\nu^{\prime}}{\nu^{\prime}_c}\int^{\infty}_{\nu^{\prime}/\nu^{\prime}_c}K_{5/3}(z)dz\right) \frac{dN_{\rm e}}{d\gamma^{\prime}_{\rm e}}d\gamma^{\prime}_{\rm e},
\end{equation}
where $\nu^{\prime}_c=3\nu^{\prime}_{\rm L}/2$ and $K_{5/3}(z)$ is the modified Bessel function. The spectra of SSC emission are calculated by the strict expressions from \cite{Blumenthal1970RvMP...42..237B}.

\begin{equation}
    \frac{dN^{\prime}}{dE^{\prime}} =\frac{{3{\sigma_{\rm T}}c}}{4} \int^{\infty}_{\gamma_m} {
    \left({\int_{}^{} {\frac{dn^{\prime}}{{d\varepsilon^{\prime} }}\frac{{d\varepsilon^{\prime} }}{\varepsilon^{\prime} }} }
    \frac{{{G(q,\Gamma_{\rm e})}}}{{{\gamma^{\prime}_{\rm e}}^2}}\right)
    \frac{{dN_{\rm e}}}{{d\gamma^{\prime}_{\rm e} }}d\gamma^{\prime}_{\rm e} } 
\end{equation}
where $q=w/\Gamma_e(1-w)$, $w=E^{\prime}/\gamma^{\prime}_{\rm e}m_{\rm e}c$.

We also consider the internal $\gamma\gamma$ absorption for VHE photons in this code. The optical depth  due to internal $\gamma\gamma$ absorption is
\begin{equation}
\label{ggabsorb}
    \tau_{\gamma\gamma,\rm int}(\varepsilon^{\prime}_{\gamma},\varepsilon^{\prime})=\int_{\varepsilon_{\rm th}}\sigma_{\gamma\gamma}(\varepsilon^{\prime}_{\gamma},\varepsilon^{\prime})\frac{R}{\Gamma}\frac{dn^{\prime}}{d\varepsilon^{\prime}}d\varepsilon^{\prime}d\Omega,
\end{equation}
where $\sigma_{\gamma\gamma}$ is the cross section, $\varepsilon_{\rm th}=2m^2_{\rm e}c^4/\varepsilon^{\prime}_{\gamma}(1-{\rm cos}\theta)$ is the threshold energy of pair production. The cross section is 
\begin{equation}
    \sigma_{\gamma\gamma}(\varepsilon^{\prime}_{\gamma},\varepsilon^{\prime})=\frac{3}{{16}}{\sigma _{\rm T}}\left( {1 - {\beta^2_{\rm cm}}} \right)\left[ {2\beta_{\rm cm} \left( {{\beta^2_{\rm cm}} - 2} \right) + \left( {3 - {\beta^4_{\rm cm}}} \right)\ln \left( {\frac{{1 + \beta_{\rm cm} }}{{1 - \beta_{\rm cm} }}} \right)} \right],
\end{equation}
where {$\beta_{\rm cm}=\sqrt{1-2m^2_{\rm e}c^4/\varepsilon^{\prime}_{\gamma}\varepsilon^{\prime}(1-{\rm cos}\theta)}$}. For an intricsic flux density $F_{\nu}$, the flux density after the internal $\gamma\gamma$ absorption is $F_{\nu}(\frac{1-e^{-\tau_{\gamma\gamma,\rm int}}}{\tau_{\gamma\gamma,\rm int}})$.

For Figure \ref{fig:spectra0}, we use analytic methods from Equation (\ref{ana1})$\&$(\ref{ana2}) to compute dynamics and electron spectra, and then obtain the flux numerically with consideration of KN effects and internal absorptions.

\section{Abosorption frequency as the break frequency}
\label{App:absorp}
In the main text we use $\nu_{\rm m}$ as the break frequency between radio (97.5GHz) and X-ray (keV). Here we discuss the possibility that $\nu_{\rm a}$ is the break frequency. The spectra of synchrotron emission are also affected by self-absorption. Frequency below absorption frequency $\nu_{\rm a}$ drops rapidly as $F_{\nu}\propto\nu^{5/2}$. If the frequency break between X-ray and radio is attributed to self-absorption, $\nu_{\rm a}$ is required to be $\nu_a\geq400$GHz at $T_0+10^5$s. The expression of $\nu_{\rm a}$ depends on the values of $\nu_{\rm m}$ and $\nu_{\rm c}$. We define $\nu_{\rm a1}$ is the absorption frequency when $\nu_{\rm a}$ is in the regime $\nu_{\rm a}<\nu_{\rm m}<\nu_{\rm c}$ . The values of $\nu_{\rm a1}$ are

\begin{equation}
\nu_{\rm a1}= \left\{
\begin{array}{ll}
     5.3{\rm GHz} E^{1/5}_{\rm iso,55}n^{3/5}_{\rm 0,-1}k^{-1}_{\rm e,-1}\epsilon ^{1/5}_{\rm B,-4},   & k=0\\
     0.09{\rm GHz} E^{-2/5}_{\rm iso,55}A^{6/5}_{\star,-1}k^{-1}_{\rm e,-1}\epsilon ^{1/5}_{\rm B,-4}t^{-3/5}_{5}. & k=2
\end{array}
\right.
\end{equation}
Apparently, $\nu_{\rm a1}$ dissatisfy the condition $\nu_{\rm a}\geq400$GHz unless we employ a very small value of $k_{\rm e}$., which are $k_{\rm e}\sim10^{-3}$ in constant medium and $k_{\rm e}\sim10^{-5}$ in wind-like medium. However, such parameters are extremely small and fail to produce sufficient flux when the isotropic energy is $10^{55}$erg. Although $\nu_{\rm a1}$ is more sensitive to the density in wind profile $\nu_{\rm a1}\propto A^{6/5}_{\star}$, large wind density $A_{\star}>1$ enhances the TeV-keV ratio at later epochs, resulting overproduction in TeV band. 

Moreover, applying such a small value of $k_{\rm e}$ would yield a very small $\nu_{\rm m}$. When $\nu_{\rm m}$ is smaller than $\nu_{\rm a}$, expressions of $\nu_{\rm a1}$ is not applicable. We define $\nu_{\rm a2}$ as the absorption frequency when $\nu_{\rm m}<\nu_{\rm a}<\nu_{\rm c}$. The absorption frequency $\nu_{\rm a2}$ evolve as $\nu_{\rm a2}\propto t^{-\frac{3p+2}{2(p+4)}}$ in constant medium and $\nu_{\rm a2}\propto t^{-\frac{3(p+2)}{2(p+4)}}$ in wind-like medium. Since $\nu_{\rm a2}$ is decreasing more rapidly with respect to time, we must have a larger $\nu_{\rm a1}$, which corresponds to more extreme parameters.

\section{Reverse Shock in a structured jet}
\label{App:Rev}
In a structured jet, the angular energy distribution is $dE/d\Omega\propto\theta^{-a}$ and the initial Lorentz factor $\Gamma_0$ may have an angular distribution $\Gamma_0\propto\theta^{\rm -k_\Gamma} $ \citep{Zhang&Wang2023}. If the $k_{\Gamma}\leq1$, the angular edges seen by observers are all decelerated since the radiative cone is $\theta\sim\Gamma^{-1}$. Hence, the forward shock dynamics in independent of the distribution of $\Gamma_0$. In this case, emissions from the forward shock can be described by Equation (\ref{forward}). 
For the wind medium, the angular profile of the Lorentz factor of the shocked shell follows 
\begin{equation}
\Gamma_3 (\theta) =\Gamma_0 (\theta) \left(\frac{R(\theta)}{R_{\rm dec}(\theta)}\right)^{-g}
\end{equation}
where $g=1$ \citep{Zou2005MNRAS.363...93Z,Gao2013NewAR..57..141G} for a  reverse shock in the thin-shell approximation and $R_{\rm dec}$ is the deceleration radius of the ejecta, which scales as $R_{\rm dec}(\theta)\propto E_{\rm\Rom{2},iso}(\theta)\Gamma_0(\theta)^{-2}$. 

Our model used $k_{\Gamma}=1$, so $\Gamma_3\propto \theta^{-1}$.  Using $R(\theta)=2\Gamma_3(\theta)^2 t$, one obtains $\Gamma_3(\theta)\propto t^{-\frac{1}{4-a}}$, $\Gamma_0(\theta)\propto t^{-\frac{1}{4-a}}$, $E_{\rm\Rom{2},iso}(\theta)\propto t^{-\frac{a}{4-a}}$.
The self-absorption frequency of reverse shock is $\nu_{\rm a,rs}\propto E^{\frac{6p-4}{7(p+4)}}_{\rm\Rom{2},iso}(\theta)\Gamma^{-\frac{24p-16}{7(p+4)}}_{0}(\theta)t^{-\frac{13p+24}{7(p+4)}}$ ($\nu_{\rm m,rs}<\nu_{\rm a,rs}$). We find $\nu_{\rm a,rs}\propto t^{-1}$ when $k_{\Gamma}=1$, which is closed to the observed scaling relation $\nu_{\rm a,rs}\propto t^{-1.08\pm0.04}$ from \cite{Bright2023NatAs...7..986B}. Correspondingly, the minimum frequency is $\nu_{\rm m,rs}\propto E^{\frac{6}{7}}_{\rm\Rom{2},iso}(\theta)\Gamma^{-\frac{24}{7}}_{0}(\theta)t^{-\frac{13}{7}}\propto t^{-1}$ and peak flux density is $F_{\rm\nu,max,rs}\propto E^{\frac{23}{21}}_{\rm\Rom{2},iso}(\theta)\Gamma^{-\frac{29}{21}}_{0}(\theta)t^{-\frac{23}{21}}\propto t^{-\frac{3}{4-a}}$, respectively.

Thus the peak flux density of self-absorption $F_{\rm\nu,a}=F_{\rm\nu,max}(\frac{\nu_{\rm a,rs}}{\nu_{\rm m,rs}})^{\frac{1-p}{2}}\propto t^{-\frac{3}{4-a}}$ is consistent with the observed values $F_{\rm\nu,a}\propto t^{-0.70\pm0.02}$ when the jet structure is flat with $a=0$. 
After the crossing of absorption frequency, the reverse shock emission declines as $F_{\nu}=F_{\rm\nu,max}(\frac{\nu}{\nu_{\rm m,rs}})^{\frac{1-p}{2}}\propto t^{-\frac{3}{4-a}-\frac{p-1}{2}}$. Assuming $p=2.2$, the slopes of decaying phase are $F_{\nu}\propto t^{-1.35}$ for $a=0$ and $F_{\nu}\propto t^{-1.54}$ for $a=0.8$.

\bibliography{reference}{}
\bibliographystyle{aasjournal}

\end{document}